\documentclass{elsart}

\newcommand{\be}{\begin{equation}}
\newcommand{\ee}{\end{equation}}
\newcommand{\ben}{\begin{eqnarray}}
\newcommand{\een}{\end{eqnarray}}

\newcommand{\tr}{{\mathrm{Tr}}}

\usepackage{amssymb}

\usepackage[pctex32]{graphics}
\begin{document}

\begin{frontmatter}

\title{Escort--Husimi distributions, Fisher information and nonextensivity}

\author[fisica]{F. Pennini},
\ead{pennini@fisica.unlp.edu.ar}
\author[fisica]{A. Plastino \corauthref{cor}}
\ead{plastino@fisica.unlp.edu.ar}

\corauth[cor]{Corresponding author.
                 Phone: +54-221-451 1412;
                 Fax:   +54-221-425 8138}

\address[fisica]{Instituto de F\'{\i}sica (IFLP)\\
                 Universidad Nacional de La Plata and \\
                 Argentina's National Research Council (CONICET)\\
                 C.C. 727,  1900 La Plata, Argentina}

\begin{abstract}

We evaluate generalized information measures constructed with
 Husimi distributions and connect them with the Wehrl entropy, on
the one hand, and with thermal uncertainty relations, on the other
one. The concept of escort distribution plays a central role in
such a study. A new interpretation concerning the meaning of the
nonextensivity index $q$ is thereby provided. A physical  lower
bound for $q$ is also established, together with a ``state
equation" for $q$ that transforms the escort-Cramer--Rao bound
into a thermal uncertainty relation.


\end{abstract}
%\maketitle
%\newpage
\end{frontmatter}
%%%%%%%%%%%%%%%%%%%%%%%%%%%%%%%%%%%%%%%%%%%%%%%%%%%%%%%%%%%%
\section{ Introduction}
%%%%%%%%%%%%%%%%%%%%%%%%%%%%%%%%%%%%%%%%%%%%%%%%%%%%%%%%%%%%%
We will be concerned here with generalizations of two important
information-theoretic uncertainty measures: those of Fisher's
($I$) \cite{Frieden,roybook}  and Wehrl's  ($W$) \cite{wehrl}. The
Wehrl entropy verifies the relation $W \ge1$ \cite{PRD2753_93},
and this bound represents
  a strengthened version of the uncertainty principle. A similar case can be made for
  $I$~\cite{Frieden,roybook}.

    In the case
  of a harmonic oscillator in a thermal state, $W$ coincides with
  the logarithmic information measure of Shannon's in the high
  temperature regime. However, it does not vanish at zero
  temperature, thus supplying a nontrivial measure of uncertainty
  due to both thermal and quantum fluctuations~\cite{PRD2753_93}.
 It has been  shown in \cite{FP03}  that intriguing  connections link $W$ to Fisher's
 information measure $I$. We will  here also generalize these
 connections.

The above referred to generalizations (indeed, nonextensive
extensions~\cite{tsallisURL,PP99}) of $W$ and $I$,
 together with their associated uncertainty applications will be
 shown to shed some light onto the {\it meaning} of the nonextensivity
 parameter $q$. Establishing adequate $q$-criteria still constitutes an open problem
 for nonextensive thermostatistics, although great progress has
 been made in deriving from first principles the appropriate $q-$value for {\it special}
   dynamical problems, some of them related to Hamiltonian systems~\cite{tsallisURL,rapisarda}.
    These systems are {\it classical} ones,
 though. Our efforts here will be directed, instead, towards
 quantum systems.

  The paper is organized as follows: in order to facilitate the reader's task,
  some preliminary material is presented in Section II.
We start our present quest in
  Section III by first generalizing the concept of Wehrl entropy to an nonextensive
  environment, obtaining a ``$q-$Wehrl'' entropy that provides us with a new interpretation
  for the index $q$ and then  studying  a Fisher's information
 measure constructed with what we call ``escort--Husimi'' distributions, which allows us
to obtain  a physical  lower bound to the nonextensivity parameter~$q$.
 Finally, some conclusions are drawn in Section IV.
%%%%%%%%%%%%%%%%%%%%%%%%%%%%%%%%%%%%%%%%%%%%%%%%%%%%%%%%
  \section{ Background material}
  %%%%%%%%%%%%%%%%%%%%%%%%%%%%%%%%%%%%%%%%%%%%%%%%%%%%%%%%
  \label{back}
\subsection{Coherent states and Wehrl information}
  In \cite{PRD2753_93} the authors discuss quantum-mechanical
  phase-space distributions expressed in terms of the celebrated
  coherent states $\vert z \rangle$ of the harmonic oscillator~
  (HO), whose Hamiltonian operator
  $\hat{H}$ is given by
      \be  \hat{H}_o= \hbar
    \omega\,\left[\hat a^{\dagger} \hat a + \frac12\right].\label{HOsHamil} \ee
The coherent states are eigenstates of the destruction operator
$\hat{a}$, i.e.,
    \be \hat a= i({2\hbar \omega
  m})^{-1/2}\hat p + (m\omega/2\hbar)^{1/2}\hat x, \label{dest} \ee
\be \label{z}    z= \frac{1}{2}\,\left(x/\sigma_x +i p/\sigma_p\right)=
  (m\omega/2\hbar)^{1/2}x + i(2\hbar \omega
  m)^{-1/2}p \equiv x'+i p',
  \ee with
  \noindent $$x'= \frac{x}{2\sigma_x};\,\,\,\, p'=
  \frac{p}{2\sigma_p};\,\,\, \sigma_x =(\hbar/2m\omega)^{1/2};\,\,\,
\sigma_p=(\hbar m \omega/2)^{1/2};\,\,\, \sigma_x \sigma_p
=\hbar/2.$$
  Variances $\sigma$ are evaluated
 for the  HO ground state. Coherent states span Hilbert's space,
 constitute an over-complete basis and obey the completeness rule
 \cite{klauder}
 \ben \label{klauder} \int \frac{\mathrm{d}^2\, z}{\pi}\,\vert z \rangle\langle z
 \vert &=& \int \frac{\mathrm{d}p\,\mathrm{d}x}{2\pi\hbar}\,\vert
 p,x
\rangle\langle p,x \vert=  1 \cr
 \mathrm{d}^2z&=&\mathrm{d}\,\Re(z)\,\,\mathrm{d}\,\Im(z)
 =\frac{\mathrm{d}p\,\mathrm{d}x}{2\hbar}\equiv \mathrm{d}p'\,\mathrm{d}x'. \een
Consider now a system characterized by a Hamiltonian $\hat{H}$.
Husimi has introduced an important  distribution function
~\cite{PRD2753_93,husimi} \be \mu(p,x)=\langle z|\hat{\rho}|z
\rangle \label{cstate}, \ee associated to the density matrix
$\hat{\rho}$ of the system. The function $\mu(p,x)$ is
  normalized in the fashion \be \label{normando} \int \frac{\mathrm{d}p\, \mathrm{d}x }{2\pi
 \hbar}\,
  \mu(p,x)= \int \frac{\mathrm{d}p'\, \mathrm{d}x'}{ \pi
 }\,
  \mu(p',x')=1, \ee
which makes it evident that $x'$ and $p'$ are the ``natural"
$\mu-$variables, of (HO-ground state) variance unity.
   The distribution $\mu$ is indeed a
  Wigner distribution smeared over a phase-space region of size
  $\hbar$ \cite{PRD2753_93}. It is important for our present
  purposes to remark  that the Husimi distribution~(HD) is {\it
 a positive definite one} \cite{PRD2753_93}. The~HD may be thought
 of as a ``classical'' distribution over
 phase-space~\cite{schnack}.

 One of the main tenets of Information Theory is that one can associate
 an information measure to any probability distribution
 \cite{katz}.   The Shannon information measure associated to the Husimi distribution is
  called the Wehrl
  entropy $W$~\cite{wehrl}
  \be
  W=-\int \frac{\mathrm{d}p\, \mathrm{d}x}{2 \pi \hbar} \mu(p,x)\, \ln
  \mu(p,x),\label{i1}.\ee
As shown by Lieb~\cite{lieb}, this special entropic form verifies
the inequality \be \label{lieb} W \ge 1.\ee
%%%%%%%%%%%%%%%%%%%%%%%%%%%%%%%%%%%%%%%%%%%%%%%%
\subsubsection{Canonical Husimi distribution}
%%%%%%%%%%%%%%%%%%%%%%%%%%%%%%%%%%%%%%%%%%%%%%%%
 Let $\hat O$ be an operator
 relevant for the system's description.
 The ``thermal'' mean value of $\hat O$  in Gibbs' canonical
ensemble is given by \cite{schnack} \ben \label{termal} \langle
\hat O \rangle &=& \tr[\hat{\rho}\, \hat O] \cr \hat{\rho}
&=&Z^{-1}e^{-\beta \hat{H}}; \,\,\, Z=\tr(e^{-\beta \hat{H}}),
\een
 with $\hat{\rho}$ the system's canonical density matrix, $Z$ the pertinent partition
function, and $\beta=1/kT$, being $T$ the temperature,
  with $k$ the Boltzmann constant, to be set equal to unity
  hereafter.
\subsubsection{HO Husimi distribution}
  In the important HO instance ($\hat{H} \equiv \hat{H}_o$),
if we denote with $|n\rangle$  the
  HO-eigenstates, associated to the eigenvalues $ E_n=\hbar
  \omega\left(n+1/2\right)$,
   one has ~\cite{PRD2753_93}
   \be \langle z|\rho|z \rangle=\frac{1}{Z} \sum_{n} e^{-\beta
  H}|\langle z|n\rangle|^2;\,\,\,
  |\langle
  z|n\rangle|^2=\frac{|z|^{2n}}{n!}e^{-|z|^2},
  \ee
entailing that  \be \label{HOhusimi}  \mu(p,x)=(1-e^{-\beta \hbar
  \omega})e^{-(1-e^{-\beta \hbar \omega})|z|^2}, \ee or, in terms of the ``natural"
  variables $p'$ and $x'$
  \be \mu(p',x')= (1-e^{-\beta \hbar
  \omega})e^{-(1-e^{-\beta \hbar \omega})[p'^2+x'^2]},  \ee
gives the HO-Husimi distribution,
    which, after integration over
  the phase space, yields an HO Wehrl's entropy  \be W(HO)=1-\ln(1-e^{-\beta \hbar
  \omega}),\label{i0} \ee
that is the Lieb's  HO-thermal uncertainty relation \cite{lieb} .
  Finally, notice that
  \ben \label{meansqerror} e^2_{\vert z \vert}(\beta,\omega)&\equiv& e^2_{\vert z
  \vert}
   =\int
  \frac{\mathrm{d}p'\mathrm{d}x'}{\pi}\,\mu(p',x')\,\vert z \vert^2 - \left[\int
  \frac{\mathrm{d}p'\mathrm{d}x'}{\pi}\,\mu(p',x')\,\vert z \vert \right]^2
  \cr
  &=& \langle \vert z \vert^2 \rangle - \langle \vert z \vert \rangle^2  = \int
  \frac{\mathrm{d}p'\mathrm{d}x'}{\pi}\,\mu(p',x')\,(p'^2+x'^2)=
     \frac{1}{1-e^{-\beta\hbar\omega}}, \een
ranges equals unity (Heisenberg's uncertainty lowest limit) for
$T=0$ and diverges  as $T\rightarrow \infty$, the typical behavior
of a {\it thermal uncertainty relation}.

\subsection{A brief primer on Fisher's Information measure}
\label{Fisher}
  A very important information measure is that advanced by  R.~A.
  Fisher in the twenties (a detailed study can be found in
  references~\cite{Frieden,roybook}). Let us consider a system that is specified by a physical
  parameter $\theta$,  while {\bf x} is a stochastic variable $({\bf x}\,\in\,\Re^{N})$
  and
  $f_\theta({\bf x})$ the probability density for ${\bf x}$,
  which depends on the parameter $\theta$.  An observer makes a
  measurement of
   ${\bf x}$ and
  has to best infer $\theta$ from this  measurement,
   calling the
    resulting estimate $\tilde \theta=\tilde \theta({\bf x})$. One
   wonders how well $\theta$ can be determined. Estimation theory~\cite{cramer}
   asserts that the best possible estimator $\tilde
   \theta({\bf x})$, after a very large number of ${\bf x}$-samples
  is examined, suffers a mean-square error $e^2$ from $\theta$ that
  obeys a relationship involving Fisher's $I$, namely, $Ie^2=1$,
  where the Fisher information measure $I$ is of the form
  \be
  I(\theta)=\int \,\mathrm{d}{\bf x}\,f_\theta({\bf
  x})\left\{\frac{\partial \ln f_\theta({\bf x})}{
  \partial \theta}\right\}^2  \label{ifisher}.
  \ee

  This ``best'' estimator is called the {\it efficient} estimator.
  Any other estimator must have a larger mean-square error. The only
  proviso to the above result is that all estimators be unbiased,
  i.e., satisfy $ \langle \tilde \theta({\bf x}) \rangle=\,\theta
  \label{unbias}$.   Thus, Fisher's information measure has a lower bound, in the sense
  that, no matter what parameter of the system  we choose to
  measure, $I$ has to be larger or equal than the inverse of the
  mean-square error associated with  the concomitant   experiment.
  This result,
  \be \label{crao} I\,e^2\,\ge \,1,\ee is referred to as the
  Cramer--Rao bound \cite{roybook}. The celebrated Uncertainty
  Principle of Heisenberg's can be shown to constitute a special
  instance of (\ref{crao}) \cite{roybook}. On account of (\ref{crao})
  one is in a position to state that $I$ provides us with a
  {\it positive} amount of information \cite{roybook}, as opposite
  to Shannon's entropy, that measures {\it ignorance} \cite{katz}.
  Also, the latter is a global measure, while $I$ is a local one
  \cite{roybook}. If $y_1,y_2,\ldots,y_n$ are  $n$ relevant
  parameters of the problem at hand (possibly including $\theta$, but not, of course,
  ${\bf x},$ that is integrated over), we will re-write (\ref{crao})
  in the fashion
\be \label{crao2} F(y_1,y_2,\ldots,y_n) \equiv I\,e^2\,\ge \,1.\ee
In the case of the harmonic oscillator, for instance, these
parameters are the inverse temperature and the frequency.

 A particular $I-$case is of great importance:
 that of translation families~\cite{roybook,Renyi},
  i.e., that in which $I$ is a functional of distribution functions (DF) whose {\it
   form} does not change under $\theta-$displacements. These DF
   are shift-invariant (\`a la Mach, no absolute origin for
  $\theta$), and for them
   Fisher's information measure (FIM) adopts the somewhat simpler appearance
  \cite{roybook}
   \be\label{shift}
  I( shift\,\,\,\,invariant)=\int \,\mathrm{d}{\bf x}\,f({\bf x})\,\left\{\frac{\partial \ln
  f({\bf x})}{
  \partial {\bf x}}\right\}^2.
   \ee
   This shift-invariant form of $I$ has encountered many physical
   applications~\cite{roybook}.
\vskip 3mm
   We will be concerned below
with a  {\it special} FIM-form called the ``escort--Fisher
measure''. It was devised, for a nonextensive setting
~\cite{tsallisURL,PP99}, by Plastino, Plastino, Miller, and
Pennini in \cite{Renyi,PPM}. Let us remind the reader first of
all of the useful concept of escort probabilities
(see~\cite{beck} and references therein). Given a normalized,
discrete (continuous) probability distribution~(PD) $P(i)$
($f({\bf x})$), its associated escort~PD of order $q$ ($q$ any
real parameter) is defined, for the discrete or continuous case
as, respectively \cite{beck}, \be \label{beck} P^{(q)}(i)=
\frac{P(i)^q }{\sum_i P(i)^q};\,\,\,or\,\,\, f^{(q)}({\bf x})=
\frac{f({\bf x})^q }{\int d{\bf x}\, f({\bf x})^q}.\ee In the
case of complex scenarios involving a PD $P(i)$ ($f({\bf x})$),
it is often the case that the associated escort~PD's yield more
insights into the concomitant dynamics than the
original~PD~\cite{beck}. The escort~FIM is then just Fisher's
measure expressed as a functional of a escort distribution of
order $q$ \cite{Renyi}

\be \label{escofi} I^{(q)}(\theta)=\int \,\mathrm{d}{\bf
x}\,f^{(q)}_\theta({\bf
  x})\left\{\frac{\partial \ln f^{(q)}_\theta({\bf x})}{
  \partial \theta}\right\}^2,\ee which obeys, instead of
  (\ref{crao}), the ``escort Cramer--Rao relation"~\cite{Renyi}

  \be \label{qrao} F_q \equiv  q^2\,I^{(q)}\,e_q^2 \ge 1,
   \ee where $e_q^2$, of course, stands for the mean-square error
      evaluated with the escort distribution (compare with Eq. (\ref{meansqerror})).

%%%%%%%%%%%%%%%%%%%%%%%%%%%%%%%%%%%%%%%%%%%%%%%%%%%%%%%%
\subsection{Nonextensive thermostatistics and escort
distributions}
%%%%%%%%%%%%%%%%%%%%%%%%%%%%%%%%%%%%%%%%%%%%%%%%%%%%%%%

Nonextensive thermostatistics is regarded by many authors as  a
new paradigm for statistical mechanics (see, for instance,
~\cite{tsallisURL,PP99,t_csf6,t_bjp29,PPpla94,TMP,ley0,Abe1,OLM}
and references therein). It is based on Tsallis' nonextensive
information measure \be S_q = -\int \mathrm{d}{\bf x}\,p({\bf
x})^q\,\ln_q p({\bf x}),\ee
 where
$p({\bf x})$ is a normalized probability density defined for ${\bf
x}\in\Re^N$ and $\ln_{q}(x)=(x^{1-q}-1)/(1-q)$ is the so-called
$q-$logarithmic function \cite{tsallisURL}, a generalization of
the standard logarithmic function.
 The real parameter $q$ is called the index of
nonextensivity, the conventional Boltzmann--Gibbs statistics being
recovered in the limit $q\rightarrow 1$.

A typical feature of nonextensive thermostatistics is that of
employing expectation values constructed with escort~PD's. This
is, if the quantity $\Delta$ takes the value $\Delta_i$ for the
event $i$ of probability $P(i)$, then \cite{Renyi} \be
\label{nosotros} \langle \Delta \rangle_q =
\sum_i\,P^{(q)}(i)\,\Delta_i, \ee is to be regarded as the
expectation value of $\Delta$ in using the MaxEnt
approach~\cite{katz} in conjunction with $S_q$ \cite{TMP,OLM}.
Tsallis' nonextensivity index is thereby {\it identified} with
the order of the underlying escort distribution,
as first pointed out in~\cite{Renyi}. %%%nuevo%%%
Summing up, current usage
of  nonextensive thermostatistics  employs three basic
ingredients: \begin{enumerate}
\item Tsallis entropy
\item MaxEnt
\item $q-$expectation values evaluated with  {\it escort
distributions}.
 \end{enumerate}
%%%%%%%%%%%%%%%end of nuevo%%%%%%%%%%%%%%%%

\label{Husimi}

%%%%%%%%%%%%%%%%%%%%%%%%%%%%%%%%%%%%%%%%%%%%%%%%%%%%%%%%
\subsection{Fisher measure and Husimi distributions}
%%%%%%%%%%%%%%%%%%%%%%%%%%%%%%%%%%%%%%%%%%%%%%%%%%%%%%%%%%

 For the reader's convenience, we summarize first of all  results obtained in~\cite{FP03} that
 involve the shift-invariant Fisher
  measure associated to the Husimi probability distribution $\mu(p,x)$.
  Firstly,  remember that
  Fisher's measure is additive~\cite{roybook}: If~ $x\,\,{\rm
  and}\,\, p$ are independent,  variables, $I(p+x)=I(p)+I(x)$,
  where we denote for $\theta\equiv\tau=(p,x)$ a point in phase-space, so that
 we face  a shift-invariance situation. One
   defines $z$ in terms of the variables $x$ and $p$, that are
  scaled by their respective variances (Cf. above the
  definition of $\vert z \rangle$). The ensuing shift-invariant Fisher
  measure is then~\cite{FP03}
  \be I(shift\,\,\,invariant)\equiv
  I_{\tau} =\int \frac{\mathrm{d}p\,\mathrm{d}x}{2
  \pi \hbar}\, \mu(p,x) \,\Gamma \label{II}\ee
  with
  \be \label{cala}
  \Gamma=\sigma_x^2\left[\frac{\partial \ln
  \mu(p,x)}{\partial x}\right]^2 +\sigma_p^2\left[\frac{\partial \ln
  \mu(p,x)}{\partial p}\right]^2,\ee
so that we can recast it in the form \be
 \Gamma =(1-e^{-\hbar\beta\omega})^2 \vert
  z\vert^2.
   \ee

   The above measure (\ref{II}) constitutes an estimation tool for {\it location
   in phase-space}.
Using here  the HO-Husimi $\mu-$expression  (\ref{cstate}),
$I_{\tau}$ adopts the appearance \be
  I_{\tau}(HO)=1-e^{-\beta \hbar \omega}\label{IF}, \ee
  so that, using (\ref{meansqerror}), we immediately verify that

\be \label{qunobnd} F(\beta,\omega) \equiv
 I_{\tau}\,e^2_{\vert z
  \vert}=1;\,\, {\rm i.e.,\,\,
Cramer-Rao's\,\, bound\,\, is \,\,reached}. \ee This result is
liable to arouse mixed feelings. On the positive side, one sees
that efficient estimation is possible at all temperatures, not
only at $T=0$. On the debit side, however, we lose  in
$F(\beta,\omega)$ all temperature-dependence in our \`a la Cramer
Fisher-estimation process. We will remedy this situation below by
recourse to escort distributions.

Finally, notice also that comparison with Eq.~(\ref{i0}) allows us
to write~ \cite{FP03}
 \be
  \label{1res} W(HO)=1- \ln{[I_{\tau}(HO)]}\Rightarrow W+\ln{[I_{\tau}]}=1,
  \ee
and we regain contextual temperature information, but using both
Wehrl's and Fisher's measures. Since the first one manages to do
this by itself (Cf. Eq. (\ref{i0})), not too much is gained. What
we really want is to obtain such a temperature context by recourse
to Fisher's information by itself, without further ado.

%%%%%%%%%%%%%%%%%%%%%%%%%%%%%%%%%%%%%%%%%%%%%%%%%%%%%%
\section{ The Husimi--Tsallis distribution}
%%%%%%%%%%%%%%%%%%%%%%%%%%%%%%%%%%%%%%%%%%%%%%%%%%%%%

%%%%%%%%%%%%%%end of nuevo%%%%%%%%%%%%%%%%%%%%%%%%
\subsection{$q-$Wehrl measure}

  We will use in
what follows the abbreviation $\tau\equiv(p,x)$ and proceed now
with the task  of generalizing Wehrl's information measure
(\ref{i1}) so as to accommodate it to a Tsallis nonextensive
environment and thus obtain the concomitant ``nonextensive-Wehrl
entropy'' $ W^{(q)}$. This is straightforwardly achieved in the
fashion

\be W^{(q)}=-\int \frac{\mathrm{d}p\,\mathrm{d}x}{2 \pi \hbar}\,
\mu(p,x)^q\,\ln_{q}\mu(p,x),\label{i3} \ee  the integration
process  encompassing the whole of phase-space.
 Explicit
evaluation of (\ref{i3}) yields, for the thermal HO, \be
 W^{(q)}(HO)=q \left\{1+\ln_{q} \left[(1-e^{-\beta \hbar
\omega})^{-1}\right]\right\}. \ee In the limit $q\rightarrow 1$ we
have the standard form obtained by Anderson {\it et al.} given by
Eq. (\ref{i0}), since $\lim_{q\rightarrow 1}\ln_{q}
\left[(1-e^{-\beta \hbar \omega})^{-1}\right]=-\ln (1-e^{-\beta
\hbar \omega})$. \vskip 3mm Note that, when the temperature goes
to zero ($\beta\rightarrow \infty$), then \be \label{interp}
W^{(q)}(HO)\rightarrow q.\ee This provides us with a {\it new
interpretation} for Tsallis' nonextensivity index $q$. It is the
the $q-$Wehrl entropy of an HO at $T=0$. Additionally, it follows
that, in a quantal regime, $q$ {\it cannot be negative}. Indeed,
according to the most basic tenet of information theory, $W^{(q)}$
represents our ignorance with regards to location in phase-space
once we know that the  probability distribution for $\tau$ is
$\mu(\tau)$ \cite{katz}. Obviously, this ignorance-amount can not
be negative. Thus, we obtain a physical lower-bound for~$q$ \be
\label{lbound} q\ge 0. \ee There is more, however. On account of
the Lieb bound $W\ge 1$ \cite{lieb}, we also get
 \be
\label{lbound2} q\ge 1. \ee

%%%%%%%%%%%%%%%%%%%%%%%%%%%%%%%%%%%%%%%%%%%%%%%%%%%%%%%%%%%%%%
\subsection{Escort--Husimi distributions}
%%%%%%%%%%%%%%%%%%%%%%%%%%%%%%%%%%%%%%%%%%%%%%%%%%%%%%%%%%%%%

%%%%%nuevo%%%%%%%
\label{nonextensive}
%\subsection{Upper bound for $q-$Husini--Fisher measures}

It is now appropriate to introduce   an
  escort $q-$Husimi distribution in the fashion
 \be
  \gamma_q(p,x)=\frac{\mu(p,x)^q}{\int \frac{\mathrm{d}p\, \mathrm{d}x}
  {2 \pi \hbar}\,\mu(p,x)^q}.
\label{escort} \ee  The associated $q-$Husimi--Fisher measure
(\ref{II}) for translation families is then \be I_\tau^{(q)}=\int
\frac{\mathrm{d}p\,\mathrm{d}x}{2 \pi \hbar}\,
\gamma_q(p,x)\,\Gamma_q=q^2\, \frac{\int
\frac{\mathrm{d}p\,\mathrm{d}x}{2 \pi \hbar}\,
\mu(p,x)^q\,\Gamma}{\int \frac{\mathrm{d}p\,\mathrm{d}x}{2 \pi
\hbar}\, \mu(p,x)^q},\label{i2} \ee since $\Gamma_q=q^2\Gamma$.
(\ref{i2}) constitutes an escort--Husimi estimation tool for {\it
location
   in phase-space}.

%%%%%%%%end of nuevo%%%%%%%%%%%

\subsection{HO application}
%\subsubsection{HO-Escort Husimi distribution}

%With reference to equation (\ref{HOhusimi}), it is important to
%point out that. in view of the Gaussian character of the
%HO-$\mu(x,p)$ PD, the following important relations hold

%\ben \label{invaq} \int \, d\tau\,\mu(x,p)\,x^2 &=& \int\,
%d\tau\,\gamma_q(x,p) x^2 \cr \int \, d\tau\,\mu(x,p)\,p^2 &=&
%\int\, d\tau\,\gamma_q(x,p) p^2 \een as it is straightforwardly
%demonstrated. This entails that  quadratic HO-thermal mean values
%are $q-$invariant. In particular, for the ``natural" variables
%$x'$ and $p'$ (Cf. Eq. (\ref{z})), all these four integrals equal
%unity.

\subsubsection{The $q$ Wehrl--Fisher connection}

  Let us connect now the measure
$W^{(q)}$ with the shift-invariant Fisher one $I_\tau$ through the
HO-Wehrl generalization \be W^{(q)}(HO)=q\,\{1-I_\tau^{q-1}\ln_q
I_\tau\} \ee where we have used the facts that i) $\ln_q
(1/x)=-x^{q-1}\ln_q x, \forall x, \forall q$, and ii) when the
parameter $q$ tends to the unity $W^{(1)}\equiv W $. In the
HO-instance we have, i)
 \be \int \frac{\mathrm{d}p\,\mathrm{d}x}{2 \pi \hbar}\,
\mu(p,x)^q=\frac{1}{q}\,(1-e^{-\beta \hbar
\omega})^{q-1},\label{muq} \ee  so that  we can check  that, in
the limit $q\rightarrow 1$, the function $\mu$ is normalized to
unity, and ii) \be \int \frac{\mathrm{d}p\,\mathrm{d}x}{2 \pi
\hbar}\, \mu(p,x)^q \vert z\vert^2=\frac{1}{q^2}\,(1-e^{-\beta
\hbar \omega})^{q-2}.\label{2muq} \ee

\subsubsection{Excitation energy}

It is opportune to recall at this point that  $\vert z \vert^2$ is
proportional to the excitation energy $E$~\cite{schnack}

\be \label{ex} E(z)= \langle z \vert \hat H \vert z \rangle -
\frac{\hbar \omega}{2} =  \hbar \omega\, \vert z \vert^2= \langle
z \vert \hbar \omega\, \hat a^{\dagger}\,\hat a \vert z \rangle.
\ee

 Several items are here to be emphasized: \begin{itemize}
 \item  from (\ref{HOsHamil}), (\ref{cala}), (\ref{i2}), and (\ref{ex})
\be I_\tau^{(q)}(HO)= q^2\,(1-e^{-\hbar\beta\omega})^2 \frac{\int
\frac{\mathrm{d}p\,\mathrm{d}x}{2 \pi \hbar}\, \mu(p,x)^q\, \vert
  z\vert^2}{\int
\frac{\mathrm{d}p\,\mathrm{d}x}{2 \pi \hbar}\, \mu(p,x)^q} \equiv
 \langle\langle \vert
  z\vert^2 \rangle\rangle_{q},\ee i.e, \be
I_\tau^{(q)}(HO)=q^2\,(1-e^{-\hbar\beta\omega})^2 \langle\langle
E/\hbar\omega\rangle\rangle_{q}, \label{IqW1}\ee which relates
the $q-$Fisher information measure to the $\gamma-$thermal mean
value of the excitation energy  (the $q-$sub-index indicates that
we are employing $q-$mean values),
\item  one finds, by inserting (\ref{muq}) and (\ref{2muq}) into  (\ref{i2}),
\be I_\tau^{(q)}(HO)=q\,\left(1-e^{-\beta \hbar \omega}\right).
\label{IqW} \ee \item  Eq. (\ref{IqW}) tells us, once again, that
$q$ cannot be negative, since it is the ratio of two
positive-definite quantities. \item Since one easily verifies that
$e_q^2$ in Eq. (\ref{qrao}) verifies $e_q^2(HO)=e^2_{\vert z
  \vert}(HO)/q^2$, the
$q-$Cramer--Rao inequality (\ref{crao}) reads here (Cf. Eq.
(\ref{qunobnd})) \be \label{newcrao} F_q(\beta,\omega)=
q^2\,I_{\tau}^{(q)}\,e_q^2= q\,I_{\tau}\,e^2\equiv q \ge
1\,\,\Rightarrow q \ge 1,  \ee as we had previously ascertained
following a Wehrl route.
 \item comparing Eq. (\ref{IqW}) with the
information (\ref{IF}) we find the following relation \be
I_\tau^{(q)}(HO)=q\,I_\tau(HO), \label{last}\ee  which connects
the escort $q-$Fisher information for
 translation families with the
 original ($q=1$) one.

 \end{itemize}

 \subsubsection{State equation for $q$}

 It has been speculated in the literature that, in some instances, one could face
 a temperature dependent $q=q(T)$ nonextensivity index (see, for instance
  \cite{OLM,casas,Abe2}, and references therein).  In such a vein, let us  assume that
the parameter $q$ is indeed a  function of $\beta$. Consideration
of the $q-$Cramer--Rao bound (\ref{qrao}) leads to the following
idea, in order to gain more insight into (\ref{newcrao}):
extremize $q\,I_{\tau}^{(q)}$ by deriving it with respect to
$\beta$ so as to obtain an equation for $q$ as a function of the
inverse temperature. Setting \be \frac{\mathrm{d}[q
I_{\tau}^{(q)}]}{\mathrm{d}\beta}=0, \ee we obtain a differential
equation for $q$ \be \label{estaes}
\frac{\mathrm{d}q}{\mathrm{d}\beta}\,(e^{\beta
\hbar\omega}-1)+q\hbar\omega=0 \ee whose solution is of the form
\be q(\beta)=(1-e^{-\beta\hbar\omega})^{-1}, \label{solving} \ee
and thus we find an escort-Cramer--Rao relation

\be \label{alfin} F_q(\beta,\omega) \equiv q  =
(1-e^{-\beta\hbar\omega})^{-1}.\ee The $q-$Cramer--Rao bound $
F_q(\beta,\omega)$ becomes (Cf. Eq. (\ref{meansqerror})) a thermal
uncertainty relation~(TUR) as that of Lieb's \cite{lieb}. In such
a sense one can then argue that, with $q=q(\beta)$ one
``optimizes" the information measure in the sense of transforming
the associated bound into a~TUR.
 The  ``equation of
state'' (\ref{estaes}) expresses the nonextensivity index $q$ in
terms of the temperature and the frequency. In particular, at zero
temperature we see that $q=1$. The present is the first concrete
example, as far as we know, of a temperature dependent $q=q(T)$
nonextensivity index.

%%%%%%%%%%%%%%%%%%%%%%%%%%%%%%%%%%%%%%%%%%%%%%%%%%%%%%%%%%
\section{Conclusions}
%%%%%%%%%%%%%%%%%%%%%%%%%%%%%%%%%%%%%%%%%%%%%%%%%%%%%%%%%%%
\label{Conclusions}

In this work, by recourse to the concept of escort-distribution,
we have performed a study of generalized information
measures~(GIM) constructed with the  Husimi distributions.
Investigating the connection of these~GIMs with i) the Wehrl
entropy and ii) thermal uncertainty relations, has allowed us to
find
\begin{itemize}
\item  two new interpretations for the nonextensivity index $q$ in terms of
\begin{enumerate} \item the $q-$Wehrl entropy and
\item  the $q-$escort Cramer--Rao bound \end{enumerate} \noindent for the case of the
thermal quantum harmonic oscillator.
\item a   lower bound for $q$, namely, $q=1$, obtained in two different ways.
\item a relation between the $q-$Fisher information measure
and  the thermal mean value of the excitation energy, valid also
for $q=1$.
\item a ``state-equation" that gives $q$ as a function of
$T,\,\,\omega$ and  transforms the $q-$Cramer--Rao bound
\cite{Renyi} $ F_q(\beta,\omega)=q^2\,I_{\tau}^{(q)}\,e_q^2 \ge 1$
into a thermal uncertainty relation.
\end{itemize}

The HO is, of course, much more than a mere example. Nowadays it
is of particular interest for the dynamics of bosonic or fermionic
atoms contained in magnetic traps~\cite{anderson,davis,bradley} as
well as for any system that exhibits an equidistant level spacing
in the vicinity of the ground state, like nuclei or Luttinger
liquids.

\end{document}